\documentstyle[preprint,eqsecnum,aps]{revtex}

\begin{document}
\baselineskip = 1\baselineskip    
\preprint{SLAC-PUB-8975}
\title{Observation of Plasma Focusing of a 28.5 GeV Positron 
Beam\footnote{Work supported by U.S. Department of Energy contract  
DE--AC03--76SF00515.}}

\author{
 { \renewcommand{\baselinestretch}{1.0} \Large \normalsize
 J.S.T.~Ng$^{1}$,
 P.~Chen$^{1}$,
 H.~Baldis$^{2}$~\cite{hab},
 P.~Bolton$^{2}$~\cite{nas},
 D.~Cline$^{3}$,
 W.~Craddock$^{1}$,
 C.~Crawford$^{4}$,
 F.J.~Decker$^{1}$, 
 C.~Field$^{1}$,
 Y.~Fukui$^{3}$,
 V.~Kumar$^{3}$,
 R.~Iverson$^{1}$,
 F.~King$^{1}$,
 R.E.~Kirby$^{1}$,
 K.~Nakajima$^{5}$, 
 R.~Noble$^{4}$~\cite{nas},
 A.~Ogata$^{6}$,
 P.~Raimondi$^{1}$,
 D.~Walz$^{1}$,
 A.W.~Weidemann$^{7}$
 }
}

\address{
\vspace{10pt}
 { \renewcommand{\baselinestretch}{1.0} \Large \normalsize
 $^{1}${\it Stanford Linear Accelerator Center, P.O. Box 4349, 
          Stanford, California, 94309} \\
 $^{2}${\it Lawrence Livermore National Laboratory, Livermore, 
	California, 94551}\\
 $^{3}${\it University of California, Los Angeles, California, 90024} \\
 $^{4}${\it Fermi National Accelerator Laboratory, Batavia, Illinois, 60510} \\
 $^{5}${\it High Energy Accelerator Research Organization, Tsukuba,
 Ibaraki 305-0801}\\
 $^{6}${\it Hiroshima University, Kagamiyama, Higashi-Hiroshima, 739-8526} \\
 $^{7}${\it Department of Physics and Astronomy, University of Tennessee, 
 Knoxville, Tennesee 37996} \\
 }
}

\date{\today}
\maketitle
\begin{center} {Submitted to Physical Review Letters}
\end{center}
\begin{abstract}
The observation of plasma focusing of a 28.5 GeV positron beam is reported. The plasma
was formed by ionizing a nitrogen jet only 3 mm thick. Simultaneous focusing in both transverse
dimensions was observed with effective focusing strengths of order Tesla per micron. 
The minimum area of the beam spot was reduced by a factor of $2.0 \pm\ 0.3$ by the plasma. The
longitudinal beam envelope was measured and compared with numerical calculations.

\end{abstract}



\vfill\eject




	The plasma lens has been proposed as a final focusing element to improve the luminosity
of future high energy electron-positron colliders~\cite{chen}. In this letter we report the observation, at
the FFTB facility~\cite{fftb_ref} at SLAC, of the focusing of a 28.5 GeV positron beam by a 
nitrogen plasma lens.

	The focusing mechanism for bunched relativistic beams, of either charge, is the collective reaction
of the plasma electrons to the relativistically foreshortened electromagnetic field of the 
bunch. Even in the time scale of a few picoseconds, plasma electrons are able to migrate towards a balance
between the superimposed beam field and the collective field caused by separating the charges of
the plasma. The primary effect is a partial neutralization of 
the electric field of the bunch. Whereas, in free space, the radial electric and toroidal magnetic fields 
of the bunch apply equal but opposite forces on the beam particles --- so that there is no net focusing or
defocusing --- this balance is lost in plasma and self-focusing occurs~\cite{chen}. 

	The collective motion of plasma electrons is characterized by the plasma wavelength, 
$\lambda_p = \sqrt{\pi/r_e n_p}$, $r_e$ being the classical electron
radius and $n_p$ the plasma density. Although this plasma current tends to cancel the beam's toroidal
magnetic field, its characteristic dimensions do not necessarily allow that to happen efficiently.
For interesting cases, the beam spot is smaller than the plasma wavelength so that the beam
toroidal field is spatially more concentrated than the plasma current and self-pinches the beam.
As plasma density rises further and its wavelength shortens, the fields overlap more completely
and the pinch is again suppressed.

	Focusing by plasma has been verified in previous experiments~\cite{ucla-kek-lbl} for low
energy (3.8 - 50 MeV) electron beams several mm in transverse size, and with relatively low density
plasmas ($\le 10^{14}$ ions per cc). Plasma focusing of positrons had not previously been demonstrated, and
this was the primary goal of the present work. We also sought to extend the beam energy into the
multi-GeV range, using plasma densities three orders of magnitude higher than before, and beam
spot cross-sectional areas six orders of magnitude smaller than before. (The experiment was
limited in principle by the carbon fibers of the measuring system, which would be destroyed by
spots of area less than $15~\mu\rm{m}^2$ for beam pulses of $1.5 \times 10^{10}$ 
positrons~\cite{clive}.)

	The plasma lens was created by the ionization of molecules in a pulsed gas jet. An 
$800~\mu\rm{s}$
jet of nitrogen gas was released at 2 Hz into the vacuum chamber by a fast acting solenoid valve,
timed so that the gas flow had stabilized when the positron beam passed through. The nozzle
diameter was 3 mm, the inlet pressure was 1000 psi for this experiment, and the gas jet opening
angle was $3^\circ$. The local beam line pressure bump was minimized by differentially 
pumping short
sections between sets of 2 mm aperture thin-foil irises. A gas density value
of $(6.5 \pm 2.5) \times 10^{18}~\rm{cm}^{-3}$
was obtained as an average of measurements made by Michelson interferometry and from
Bremsstrahlung observed during data taking.

	As the positron beam passed through, ionization of the jet occurred by the normal energy
loss process~\cite{Zarubin}, enhanced by avalanche collisions during acceleration by the beam field. We
have found that an improved focusing effect could be obtained by irradiating the gas with
focused, pulsed laser light. A commercially available Q-switched Nd:YAG laser delivered 10 Hz
of 10 ns pulses at wavelength 1064 nm and pulse energies near 1.2 J. For comparison, the
positron pulse width was 4 ps FWHM. The light, incident on the gas jet at right 
angles to the positron
beam direction, was line focused astigmatically to a spot of $50~\mu$m vertically by $330~\mu$m
horizontally FWHM. It is estimated that about 0.1$\%$ of the laser energy was absorbed through
collisional ionization seeded by multiphoton absorption. The superheated gas expanded in a
shock-wave shell, and thus a wide range of available plasma densities evolved over time. The
time between the laser pulse and the beam was optimized (at about 400 ns) by observing the
strength of the synchrotron radiation from the focusing. At this time the 
ionization caused by the
positron beam was much stronger than that remaining from the laser, which had largely
recombined and been spatially diluted. However, this process requires further investigation and
will not be discussed here.

\begin{figure}
\caption{Schematic layout of the SLAC Plasma Lens experiment at the FFTB. ``Final Quads" are
conventional focusing quadrupole magnets. The positron beam is deflected towards the dump by
the magnetic dipole}
\label{beamline}
\includegraphics{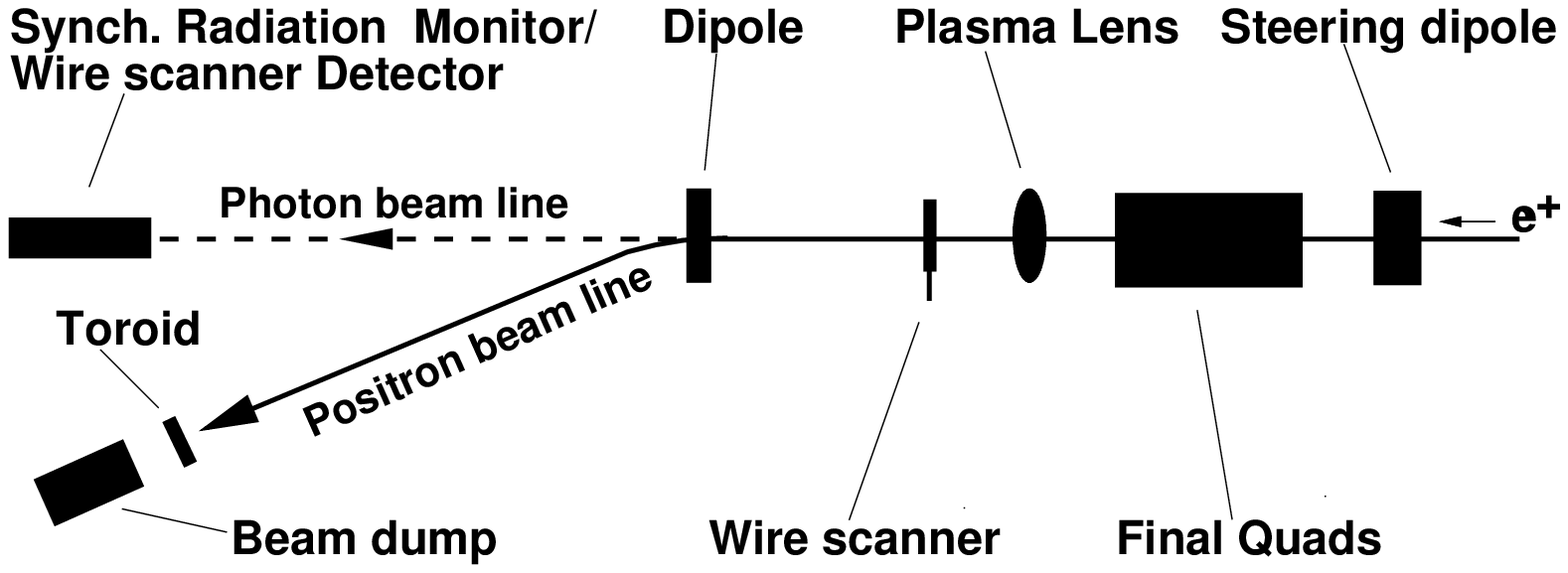}
\end{figure}

	The layout of the experiment around the plasma lens is shown in Figure~\ref{beamline}. The direction
of the positron beam defines the z-axis of the right-handed coordinate system, and y is vertical.
The beam energy was set to 28.5 GeV, with 0.6 mm long bunches, normalized emittances of
$5 \times 10^{-5}$ m-rad (x) and $5 \times 10^{-6}$ m-rad (y), and $1.5 \times 10^{10}$ particles per bunch.

	Conventional magnets were set to focus the positrons near the plasma lens, where the beam size was 
measured using a wires canner system. A carbon fiber of 7~$\mu$m diameter was moved to a known 
position between 8 and
32 mm downstream of the entrance to the gas jet. In an automated procedure, the positron beam
was steered orthogonally across the fiber in 1~$\mu$m steps (whose scale was calibrated to better
than  $\pm 5 \%$). Bremsstrahlung from the carbon traveled 33 m downstream, escaped the vacuum
pipe, showered in a 4 radiation length stack of polyethylene plates, and was detected by an air
Cherenkov counter~\cite{clive}. The correlation between steering and the Bremsstrahlung signal gave the 
beam profile. We found that the elliptical beam spot size at the lens was typically
11.5 $\mu$m  by 2.5 $\mu$m. The beam density within the $2\sigma$ contour was $2 \times 
10^{16}~\rm{cm}^{-3}$.

	Interleaved among the plates were planar 
ion chambers used to monitor the several-MeV critical energy of the synchrotron radiation emitted from the 
plasma lens (the Synchrotron Radiation Monitor or SRM). The ion chambers also detected the Bremsstrahlung 
and were used as a check on the spot size measurement.

\begin{figure}
\caption{Wire scan beam profile measurements, showing data taken
a) with no nitrogen present and b) with focusing by the nitrogen plasma. The superimposed curves 
represent Gaussian function fits.}
\label{run5-scan}
\includegraphics{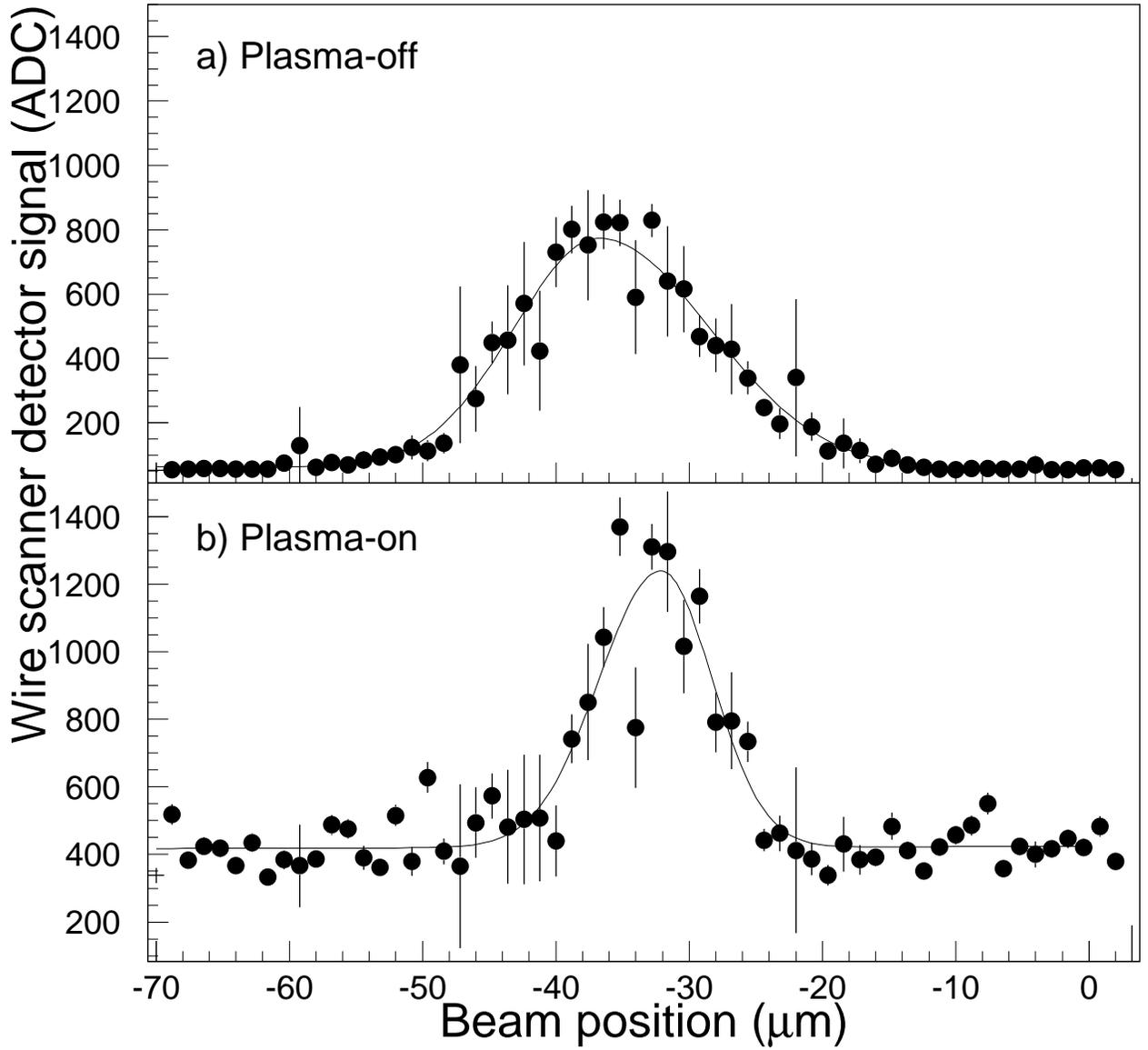}
\end{figure}

	As a way of reducing the effects of drifts, measurements of plasma-on and plasma-off
beam spot profiles were interleaved within the scan. At each scan step of the beam, signals from
four pulses were recorded with the gas jet off, and averaged,  and one was taken with the jet
firing. These were plotted, as in Figure~\ref{run5-scan}, against the scan position.

	There are several other systematic effects in the beam size measurement process. First of
all, the carbon fiber diameter contributes 1.7 $\mu$m in quadrature to the RMS beam size; this is
corrected for. Since it takes many beam pulses to make a profile measurement, shot-to-shot
fluctuations in the beam centroid position also contribute to the measured beam size. By
measuring the RMS fluctuation at a fixed fiber location and fixed beam steering, this contribution
by itself is estimated to
be 25$\%$ of the measured width, and when subtracted in quadrature, it reduces the apparent beam
width by 3$\%$. For the data being reported here, the elliptical positron beam spot also had a roll
angle of 13$^{\circ}$  with respect to the x direction, and this was accounted for.

	An important correction was necessary because the strong focusing of the plasma
increased the beam divergence at the wire scanner. As the focused beam was scanned, a portion
of the Bremsstrahlung photon cone was occluded by a downstream fixed aperture. The total
photon flux, as given by the area under the scan profile curve, was reduced by 30\%. The
systematic effect was to make the scan profile asymmetric because of the correlation between the
positron angle and position at the fiber. A check with the beam toroids confirmed that there was
no loss in the charged beam flux as it passed through the lens to the dump.

	To account for this effect, the measured beam profiles have been fitted to a Gaussian
function with different widths allowed on either side of the peak. The larger width parameter was
taken as the estimate of the unoccluded width. A first order polynomial took account of the
background accompanying the beam, the synchrotron radiation, and beam-gas Bremsstrahlung
from the jet.

	The accuracy of this procedure for determining the beam width has been estimated using
a ray-tracing simulation of the experiment. The incoming beam envelope was parameterized with
the help of upstream wire scanners, as well as jet-off local measurements at a range of z-locations.
These parameters were reproduced in the linear optics in the simulation. A model for
the beam-plasma collective focusing field was then constructed~\cite{bassetti}. This accounted for the
spatial variation of the plasma focusing strength within the bunch. The focused positrons were
followed to the carbon fiber, and the Bremsstrahlung photons traced from there to occluding
apertures downstream. The geometric parameters were adjusted to reproduce the observed loss in
the photon transmission. Using the same asymmetric Gaussian fit procedure as for the real data,
the simulation showed some variations over the z-range that lead us to assign an uncertainty of
$\pm5\%$ for data without the plasma, and $\pm15\%$ for the heavily occluded cases with plasma. These
are taken as systematic uncertainties in the beam size measurements.

\begin{figure}
\caption{Measured beam envelope Gaussian widths in the $x$- and $y$-dimensions, with and 
without plasma focusing. Inner error bars indicate the statistical uncertainty, and outer ones the 
quadrature sum of statistical and systematic uncertainties. The curves represent the
particle-in-cell simulations.}
\label{positrons}
\includegraphics{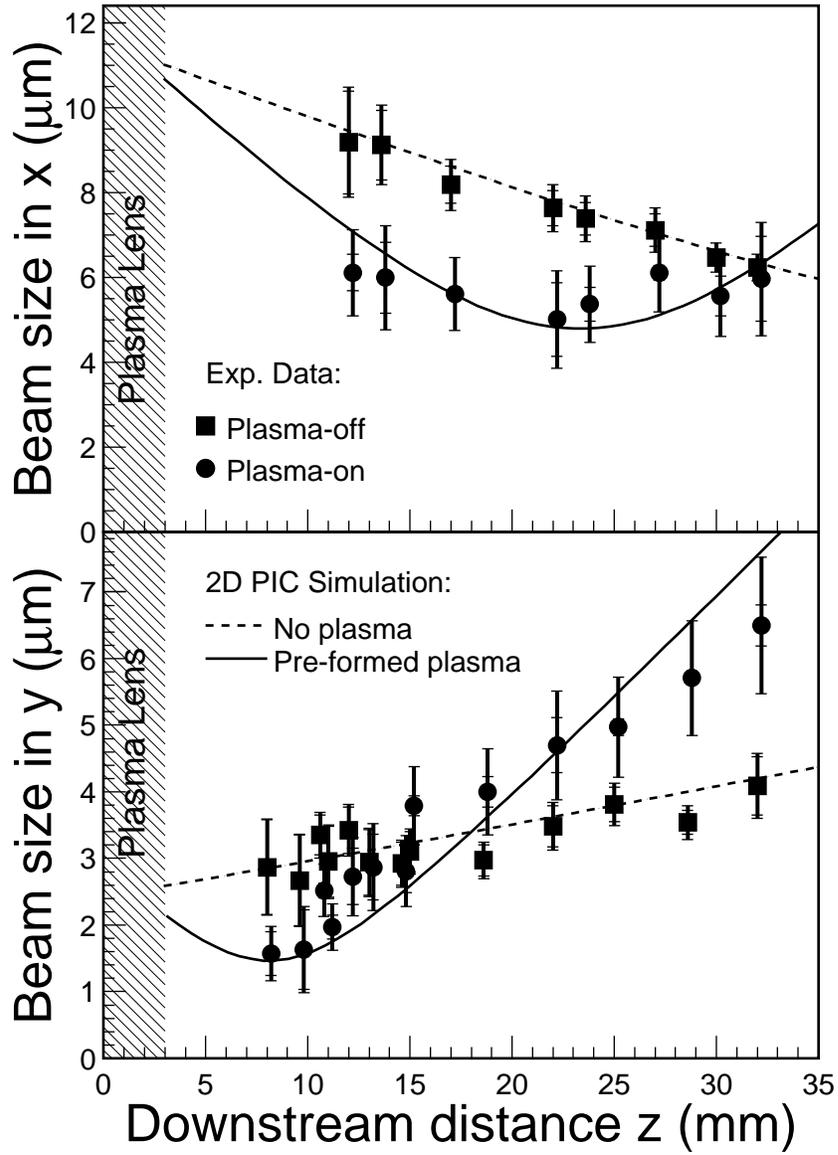}
\end{figure}

	The plasma focusing was studied by making multiple measurements of the x- and y-profiles
of the beam envelope at several values of z within the accessible range. Data at the same
settings were averaged. The plasma-on data are compared with the simultaneous plasma-off 
envelope in Figure~\ref{positrons}. The pinch is quite striking as seen in the y direction.
From the y divergence angles,
the lens quadrupole strength may be estimated to be 4 T/$\mu$m, and from the fitted waist 
positions
the effective focal length is 1.6 mm. The values in the x direction are 0.7 T/$\mu$m and 34 mm.

	Numerical integration over the expected transverse field profile of an ideal plasma lens 
shows that these values of quadrupole strength would yield synchrotron radiation with 
effective Critical 
Energies E$_c$~\cite{Sands} of 4.6 and 3.4 MeV respectively. As a test of this, for a few 
representative scans we have examined the 
signals from the SRM chambers as a function of depth in the polyethylene. A simple simulation, 
based on the electromagnetic shower code EGS4~\cite{EGS4}, was 
made of the response of the detectors to synchrotron radiation. The depth profile led to a value 
$\rm{E}_c = 4.4\pm0.3$ MeV, in satisfactory agreement with the range of expected values. 

	Luminosity depends on the inverse of the minimum cross-sectional area of the beam, and
this was reduced by a factor of 2.0 $\pm$ 0.3 by the plasma lens. However, the beam time 
available
was inadequate for tuning the x- and y-focal spots to the same z-location, either for the plasma
on or off, and so we cannot report directly the reduction of spots as they would be optimized for
luminosity. Also, aberrations are expected in plasma lenses from imperfect charge
distributions~\cite{rosen-chen}, and tuning the beam optics to minimize their
effects was not practical in this run.

	A three-dimensional particle-in-cell (PIC) electromagnetic simulation code is being
developed to describe the focusing process in this experiment for both positrons and electrons. In
particular it will allow for the $\sigma_x/\sigma_y = 4.8$ aspect ratio and the beam-induced 
ionization. At present, results are available from a
simplified two-dimensional PIC simulation. Plasma focusing in the x- and y-planes was
simulated separately using a round beam configuration where the size was
selected to be $\sigma_x$ or $\sigma_y$
as appropriate. In order to maintain the same focusing strength, the number of beam particles was
scaled by the aspect ratio between the two cases. A uniform plasma was used, 3 mm thick with a
density of $5\times 10^{17}$ cm$^{-3}$. Since this density was greater than that of the 
beam, the focusing
effect was in the self-limiting range and insensitive to the exact density assumed. Considering the
simplifications that have been made, the results, shown as lines in Figure~\ref{positrons}, agree reasonably
with the data, indicating that more complete simulations will have predictive power.

	In summary, we have reported the observation of focusing of high energy positron beams
by a thin lens of dense plasma. Simultaneous reduction of the beam spot in both dimensions was
observed. Future work toward applying the technique to linear colliders will, however, require
improvements to the available simulation codes and to the understanding of plasma production
processes.

	We wish to thank the support staff at the various institutions for their hard work in
the design and construction of the apparatus, in particular G. Collet, K. Krauter, G. Mazaheri, R.
Rogers, Y. Sung and W. Kaminskas (deceased) and his staff. The experiment would not have
been possible without the dedicated efforts of the SLAC accelerator operations staff.
This work is supported in part by the 
Department of Energy under contracts 
DE-AC02-76CH03000, DE-AC03-76SF00515, DE-FG03-92ER40695,
and DE-FG05-91ER40627, and the Lawrence Livermore National Laboratory
through the Institute for Laser Science and Applications,
under contract No. W-7405-Eng-48; and by the 
US-Japan Program for Cooperation in High Energy Physics.

\newpage

\newpage

\newpage

\end{document}